\documentclass[aps,prd,groupedaddress,showpacs,nofootinbib]{revtex4}
\usepackage{graphicx,amssymb,amsmath,epsfig}
\usepackage[english]{babel}
\def\comment#1{}
\newcommand{\be}{\begin{equation}}\newcommand{\ee}{\end{equation}}
\newcommand{\bea}{\begin{eqnarray}}\newcommand{\eea}{\end{eqnarray}}
\newcommand{\beaa}{\begin{eqnarray}}\newcommand{\eeaa}{\end{eqnarray}}
\newcommand{\ba}{\begin{array}}\newcommand{\ea}{\end{array}}
\newcommand{\bit}{\begin{itemize}}\newcommand{\eit}{\end{itemize}}
\newcommand{\ben}{\begin{enumerate}}\newcommand{\een}{\end{enumerate}}

 \newcommand{\sfrac}[2]{\raisebox{0.095ex}{\scriptsize${\frac{#1}{#2}}$}}

\begin{document}

\title{Equivalence Principle and Field Quantization in Curved Spacetime
}

\author{H. Kleinert}

%\vspace{2mm}

       \address{
Institut f\"ur Theoretische Physik,
 Freie Universit\"at Berlin,
Arnimallee 14, D14195 Berlin, Germany\\
and\\
ICRANeT, Piazzale della Republica 1, 10 -65122, Pescara, Italy
}

%\maketitle

\vspace{2mm}

\begin{abstract}

 To comply with the equivalence principle,
fields in curved spacetime
can be quantized only
in the neighborhood of each point, where one can construct
a freely falling {\em Minkowski\/} frame with {\em zero\/} curvature.
In each such frame,
the geometric forces
of gravity can be replaced by
a selfinteracting spin-2
field, as proposed by Feynman in 1962.
 At {\em any fixed\/} distance $R$ from a black hole,
the vacuum in
each freely falling volume element
acts like a thermal
bath of all particles with Unruh temperature $T_U=\hbar GM/2\pi c R^2$.
%%%%%%new %%%%%%%%%
At the horizon $R=2GM/c^2$, the falling vacua show
the Hawking temperature
$T_H=\hbar c^3/8\pi GMk_B$.

%%%%%%%%%%%%%%%%%%%%%%%

\end{abstract}


\maketitle

1.) When including Dirac fermions into the
theory of gravity,
it is important to remember
that the Dirac field is initially defined only
by its transformation behavior under the Lorentz group.
The invariance under general coordinate transformations
can be incorporated only with the help
of a vierbein field $e^ \alpha {}_\mu(x)$
which couples Einstein indices with Lorentz indices $ \alpha $.
These serve to define
anholonomic coordinate differentials $dx^ \alpha $ in
curved spacetime $x^\mu$:
\begin{equation}
dx^ \alpha =e^ \alpha {}_\mu(x)dx^\mu,
\label{@}\end{equation}
which at any given point have a Minkowski metric:
\begin{equation}
ds^2 = \eta _{ \alpha  \beta }dx^ \alpha dx^ \beta ,~~~~
\eta _{ \alpha  \beta }=
\left(
\begin{array}{cccc}
1&0&0&0\\
0&-1&0&0\\
0&0&-1&0\\
0&0&0&-1
\end{array}
\right)
=\eta ^{ \alpha  \beta }.
\label{@MIN}\end{equation}
With the help of the vierbein field one can write the action
simply as \cite{MVF}
\begin{equation}
{\cal A}=\int d^4x  \sqrt{-g} \bar\psi(x)[
 \gamma ^ \alpha e_{ \alpha }{}^\mu(x)(i\partial _\mu-
\Gamma_{\mu}^{ \alpha  \beta }
\sfrac{1}{2}\Sigma_{ \alpha  \beta })
 -m]\psi(x), ~~
~g_{\mu \nu }(x)\equiv
 \eta _{ \alpha  \beta }e^ \alpha {}_\mu   (x)
e^ \beta  {}_\nu(x) ,~~~~
\label{@LL}\end{equation}
where $\Sigma_{ \alpha  \beta }$ is the spin matrix, which is formed from the commutator
of two Dirac matrices as
$i[\gamma_ \alpha , \gamma _ \beta ]/4$, and
$\Gamma_{\mu}^{ \alpha  \beta }
\equiv
e^ \alpha {}_ \nu
e^ \beta  {}_  \lambda
\Gamma_{\mu}{}^{ \nu  \lambda  }$ is the spin  connection,
It is constructed
from
 combination of the so-called objects of anholonomity
$
\Omega_{\mu \nu }{}^ \lambda
 =\sfrac{1}{2}[
 e_\alpha {}^ \lambda \partial _\mu e^ \alpha {}_ \nu
-(\mu\leftrightarrow \nu )]  $, by
taking
the sum
$
\Omega^{\mu \nu  \lambda}-
\Omega^{\nu  \lambda\mu}+
\Omega^{  \lambda\mu \nu }$
and lowering
two indices with the help of the metric
$g_{\mu \nu }(x)$.

The theory of quantum fields
in curvilinear spacetime has
been set up on the basis of this Lagrangian, or a simpler version for bosons
which we do not write down.
The classical field equation is solved on the background metric
$g_{\mu \nu }(x)$
in the entire spacetime. The field is expanded into
the solutions, and the coefficients are quantized
by canonical commutation rules, after which they serve as
creation and annihilation operators on some global vacuum
of the quantum system.

The purpose of this note is to make this
this procedure
compatible with the equivalence principle.
%%%% 1st correction %%%%%%%%%%one of the
%%%%%%%%%%holiest shrines of theoretical and experimental physics.
~~\\

2.) If one wants to quantize the theory in
accordance with
the equivalence principle
one must introduce
creation and annihilation operators of proper elementary particles.
These, however,
are defined as irreducible representations of the
Poincar\'e group with a given mass and spin.
The symmetry transformations of the Poincar\'e group
can be performed
only in a Minkowski spacetime.
According to Einstein's theory, and confirmed by Satellite experiment,
we can remove gravitational forces locally at one point.
The neighborhood will still be subject to gravitational fields.
For the definition of elementary particles
we need only a small neighborhood.
In it,
the geometric forces
can be replaced
by the forces
coming from the spin-2 gauge field theory of gravitation,
which was developed by R. Feynman
in his~1962 lectures at Caltech \cite{Feynmann1}.
This can be rederived
by
expanding of the metric in
powers its deviations from the
flat Minkowski metric.
We define a Minkowski frame $x^a$
around the
point of zero gravity, and extend it
to an entire finite box without spacetime curvature.
Inside this box,
particle experiments can be performed
and the transformation properties under the
Poincar\'e group can be identified.

Inside the box, the
fields are governed
by the flat-spacetime action
\begin{equation}
{\cal A}=\int d^4x \sqrt{-g} \bar\psi(x)\{ \gamma ^ a
 e_{ a }{}^b
 (i\partial _ b-
\Gamma_{a}^{ bc }
\sfrac{1}{2}\Sigma_{ bc })
 -m\}\psi(x).
\label{@LL2}\end{equation}
In this expression,
 $ e_{ a }{}^b= \delta_a{}^b+\sfrac{1}{2}h _{ a }{}^ b(x)$.
The metric and
the spin connection
are defined as above,
exchanging
the indices $ \alpha , \beta ,\dots$
by $a,b,\dots~$.
All quantities must be expanded in powers of
 $h_a{}^b$.

Thus we have arrive at a standard local field theory
in the freely falling Minkowski laboratory
around the
point of zero gravity.
This action is perfectly Lorentz invariant,
and the Dirac field can now be quantized without problems,
producing an irreducible representation
of the Poincar\'e group
with states of definite momenta and spin orientation
$|{\bf p},s_3 [m,s]\rangle$.

The Lagrangian governing
the dynamics of the
field
$h_ a {}^ b (x)$
%-from-$ \eta_ a {}^ b $-field $h_ a {}^ b (x)$
is well known from Feynman's lecture
\cite{Feynmann1}. If the laboratory is sufficiently small,
we may
work with the Newton approximation:
\begin{equation}
{\cal A}^h=-\frac{1}{8 \kappa }\int d^4x
\,h_{ab}
 \epsilon ^{cade}
 \epsilon _{c}{}^{bfg} \partial _d \partial _fh_{eg}+\dots~
, ~~~ \kappa = {8\pi G/c^3}, ~~G={\rm Newton~ constant},
\label{@}\end{equation}
where $ \epsilon ^{cade}$ is the antisymmetric unit tensor.
If the laboratory is larger, for instance, if it
 contains the orbit of the planet mercury, we must include
also the first post-Newtonian corrections.

Thus, although the Feynman spin-2 theory is certainly  not a valid replacement
of general relativity, it is so in a neighborhood of
any freely falling point.

The vacuum of the Dirac field is, of course,
not universal. Each point $x^\mu$ has its own vacuum
state restricted to the
associated freely falling Minkowski frame.
~\\

3.) There is an immediate consequence of this quantum theory.
If we consider a Dirac field in a black hole,
and go to the neighborhood of any point,
the quantization has to be performed in the freely falling Minkowski frame
with smooth forces.
These are incapable of creating pairs.
%%%%%%%%2nd correction %%%%%%%%%%%%from here to below %%%%%%%%%%%%%%%%%%%%%
An observer at a fixed distance $R$
from the center, however,
sees the vacua of these
Minkowski frames pass by with  acceleration $a=GM/R^2$,
where $G$ is Newton's constant.
At a given $R$,
the frequency factor
$e^{i \omega t}$
associated with the zero-point oscillations
of each scalar particle wave
of the world
will be Doppler shifted
to $e^{i \omega e^{i at/c}c/a}$, and this wave has frequencies
distributed with a probabilty
that behaves like $1/(e^{ 2\pi \Omega c/a}-1)$. Indeed, if we Fourier analyze
this wave \cite{SIM}:
\begin{eqnarray}
\int_\infty^\infty  dt \,e^{i \Omega t}
e^{i \omega e^{i at/c}c/a}=
e^{-\pi c/2a}
\Gamma(i \Omega c/a)e^{-i \Omega c/a \log(\omega c/a)}(c/a).
\label{@}\end{eqnarray}
we see
that  the probability to find the frequency $ \Omega $
is $ |e^{-\pi c/2a}
\Gamma(i \Omega c/a)c/a|^2$, which
is equal to $2\pi c/ (\Omega a)$ times $1/(e^{2\pi  \Omega c/a}-1)$.
The latter
is
a
thermal
Bose-Einstein distribution
 with an Unruh temperature $T_U=
 \hbar a/2\pi c k_B$
\cite{UNR}, where $k_B$ is the Boltzmann constant.
The particles in this heat bath
can be detected
by suitable  particle
reactions as described
in Ref.~\cite{DETE}.

The Hawking temperature $T_H$
is equal to the Unruh temperature
of the freely falling
Minkowski vacua
at the surface of the black hole,
which
lies at
the
horizon
$R=
r_S\equiv 2GM/c^2$.
There the
Unruh temperature is
equal to
$
T_U|_{a=GM/R^2,R=2GM/c^2}=\hbar c^3/8\pi GMk_B= T_H$.

Note that there exists a thermal bath
of nonzero Unruh temperature
$T_U(R)=T_H r_S^2/R^2$
at {\em any\/} distance $R$ from the center --- even on the
surface of the earth, where the temperature is
too small to be measurable.
In the light of this it is surprising that
the derivation of
the thermal bath from semiclassical pair creation
is based on a
coordinate singularity at the horizon \cite{PW}.

For decreasing $R$ inside the horizon, the temperature rises
to infinity, but this
radiation cannot
reach
any outside oberver.

%%%%%%%%%%%%%%%%%%%%%%%%%%%%%%%%%%%%%%%%%%%%%%%

~\\
Acknowledgment:\\[1mm]The author
thanks V. Belinski
pointing out the many papers
on the semiclassical explanation of pair creation
in a black hole.


\begin{thebibliography}{99}


\bibitem{MVF}
Our notation follows\\
{H.~Kleinert},
      {\em Multivalued Fields in Condensed Matter,
Electromagnetism, and Gravitation\/},
	  World Scientific, Singapore 2009, pp. 1--497
({\tt http://www.physik.fu-berlin.de/\~{}klei\-nert/b11}).\\
The only exception is
that the vierbein field is here called $e^ \alpha {}_ \mu$
rather than
$h^ \alpha {}_\mu$ to have the notation $h^a{}_b$ free for
the small deviations of $e^a {}_b$ from the flat limit $ \delta^a {}_b $.


\bibitem{Feynmann1}
  R.P. Feynman,
F.B. Moringo, D. Pines,
{\em Feynman Lectures on Gravitation\/} (held in 1962 at Caltech),
Frontiers in Physics, New York. 1962.
%See also Ref.~\cite{Weinberg12}.

\bibitem{SIM}
 P.M. Alsing and P.W. Milonni,
Am. J. Phys. {\bf 72}, 1524 (2004)
(arXiv:quant-ph/0401170).

\bibitem{UNR}
W.G. Unruh, Phys. Rev. D {\bf 14}, 870 (1976).

\bibitem{DETE}
A. Higuchi, G. E. A. Matsas, and D. Sudarsky, Phys. Rev. D
45, R3308 (1992); 46, 3450 (1992);
A. Higuchi, G.E.A. Matsas, D. Sudarsky,
Phys. Rev. D {\bf 45} R3308 (1992); {\it ibid.}
 {\bf 46}, 3450 (1992);
  D.A.T.
  Vanzella  and
G.E.A. Matsas,
Phys. Rev. D  {\bf 63}, 014010 (2001).
J. Mod.Phys. D {\bf 11}, 1573 (2002).

\comment{\bibitem{Weinberg12}
  S. Weinberg,
Phys. Rev. 135, B1049 (1964);
Phys. Rev. 140, B516 (1965).
}

\bibitem{PW}
M.K. Parikh, F. Wilczek,
Phys. Rev. Lett. {\bf 85}, 5042 (2000).

\end{thebibliography}
\end{document}